\documentclass[article,epsfig,graphicx,nofootinbib,showpacs,eqsecnum,aps,amsmath,mathrsfs,eufrak,amssymb,floatfix,twocolumn]{revtex4}
\usepackage{graphicx,graphics}
\usepackage{epsfig}
\begin{document}

\title{Quartic oscillator potential in the $\gamma$-rigid regime of the collective geometrical model}
\author{R. Budaca}

\affiliation{Horia Hulubei National Institute of Physics and Nuclear Engineering, RO-077125 Bucharest-Magurele, Romania}
\date{\today}
\begin{abstract}
A prolate $\gamma$-rigid version of the Bohr-Mottelson Hamiltonian with a quartic anharmonic oscillator potential in $\beta$ collective shape variable is used to describe the spectra for a variety of vibrational-like nuclei. Speculating the exact separation between the two Euler angles and the $\beta$ variable, one arrives to a differential Schr\"{o}dinger equation with a quartic anharmonic oscillator potential and a centrifugal-like barrier. The corresponding eigenvalue is approximated by an analytical formula depending only on a single parameter up to an overall scaling factor. The applicability of the model is discussed in connection to the existence interval of the free parameter which is limited by the accuracy of the approximation and by comparison to the predictions of the related $X(3)$ and $X(3)$-$\beta^{2}$ models. The model is applied to qualitatively describe the spectra for nine nuclei which exhibit near vibrational features.
\end{abstract}
\maketitle

\renewcommand{\theequation}{1.\arabic{equation}}
\section{Introduction}
\label{intro}
Shape phase transitions between different dynamical symmetries became a more interesting topic since the introduction of critical point symmetries \cite{X5,E5} which allowed a parameter free description not only of the extremes but also of their critical point. However these are not true dynamical symmetries in the sense of group reduction defined in the framework of the interacting boson model (IBM) \cite{IBM}, but fitting descriptions provided by similarly simple shapes of the potential surface in the collective geometrical model \cite{BM1,BM2}. Presently there are known two such critical symmetries, associated to the phase transitions from the spherical vibrator shape phase described by the $U(5)$ \cite{B} dynamical symmetry to the $O(6)$ \cite{WiJe} dynamical symmetry characterizing the $\gamma$-unstable nuclei and respectively to the $SU(3)$ dynamical symmetry of axial rotors. Iachello realized that the critical point potential for both transitions can be fairly well approximated by a square well potential, the resulted models being called $E(5)$ \cite{E5} and $X(5)$ \cite{X5} respectively. However a consistent algebraic treatment of the most general (up to two-body terms) IBM Hamiltonian at the critical point of the $U(5)\leftrightarrow O(6)$ transition showed results equivalent to those obtained in the geometric model with a pure quartic oscillator potential in $\beta$ shape variable identified as $E(5)$-$\beta^{4}$ model \cite{Arias,Vorov}. This particular transition have a special allure due to the exact separation of the shape variables which is not found in the other transition and consequently in the $X(5)$ model.

Imposing a certain value for the $\gamma$ shape variable, one reaches the $\gamma$-rigid version of the collective model which is interesting by itself due to its description of the basic rotation-vibration coupling \cite{Dav}. A $\gamma$-rigid version of the critical symmetry $X(5)$, called $X(3)$ was proposed not long ago \cite{X3}, revealing a similarity in the $\beta$ excited bands between the two model predictions and which is based on the exact separation of shape and angular variables. Given the openness of the question regarding the shape of the critical point potential, it is then interesting to study the $\gamma$-rigid realization of the general quartic oscillator potential with an emphasis on the pure quartic oscillator, and compare it to the models defined in the same space of variables, i.e. $X(3)$ and $X(3)$-$\beta^{2}$. It must be mentioned that such a program was never realized primarily due to the fact that quartic oscillator Schr\"{o}dinger equation cannot be analytically solved, and the existing predictions of models involving mainly the pure quartic oscillator potential are based on numerical integration. The study of the solutions corresponding to the quartic oscillator potential is very important enforced also by the fact that the next leading anharmonic term of the $\beta$ potential is $\beta^{4}$ not only in the $\gamma$-rigid case, but also in the $\gamma$ stable and unstable models, with the potential separation approximation \cite{WiJe} in the former case. The next order anharmonic term lead to a sextic potential which allows a nonvanishing minimum, being thus suitable for the description of deformed nuclei \cite{Levai,Rabug1} and even of the candidates for the $X(5)$ critical point symmetry \cite{Rabug2}. Moreover, the differential equation for $\beta$ with sextic potential is quasi-exactly solvable. It is worth to mention that treating semiclassically a second \cite{second}, fourth \cite{fourth} and sixth \cite{sixth} order quadrupole boson Hamiltonian leads to a Schr\"{o}dinger equation with a harmonic oscillator, quartic and respectively sextic potentials in a radial-like variable with a centrifugal barrier.

In this paper one investigates the $\gamma$-rigid problem for a general quartic anharmonic oscillator potential (QAOP), i.e. with both quadratic and quartic terms, by using an analytical formula for the energies of the ground and $\beta$ excited bands. Due to the scaling property of the QAOP, the energies are function only on one parameter up to an overall multiplying factor, even though the potential is defined by two parameters. The applicability of the energy formula is studied with regard to the allowed values for the ratio between the first two excited states $R_{4/2}$ taking also into account the validity of the adopted approximation. Theoretical calculations are carried out for nine nuclei exhibiting vibrational-like behaviour, the results being compared with the corresponding experimental spectra.

\renewcommand{\theequation}{2.\arabic{equation}}
\section{$\gamma$-rigid realization of the Bohr-Mottelson Hamiltonian}
\label{sec:1}
The quadrupole deformation is the fundamental mode of deformation for a spherical system. It can be described by a set of five amplitudes that form the components of a spherical tensor. The five tensorial coordinates are usually transformed by Bohr-Mottelson parametrization to three orientation angles $\Omega=(\theta_{1},\theta_{2},\theta_{3})$ and two shape variables $\beta$ and $\gamma$. In this parametrization, the potential energy of the nuclear deformation depends only on the shape variables and its general form up to the sixth order is given in the generalized collective model \cite{GnGr} as
\begin{eqnarray}
V(\beta,\gamma)&=&c_{1}\beta^{2}+c_{2}\beta^{3}\cos{3\gamma}+c_{3}\beta^{4}\nonumber\\
&&+c_{4}\beta^{5}\cos{3\gamma}+c_{5}\beta^{6}\left(\cos{3\gamma}\right)^{2}+c_{6}\beta^{6}.
\end{eqnarray}
While the corresponding kinetic energy is quadratic in the time derivatives of all variables and can be separated into a vibrational part
\begin{equation}
T_{vibr}=\frac{B}{2}\left(\dot{\beta}^{2}+\beta^{2}\dot{\gamma}^{2}\right),
\end{equation}
and a rotational part given by
\begin{equation}
T_{rot}=\frac{1}{2}\sum_{k=1}^{3}\omega_{k}^{2}\mathcal{I}_{k},
\end{equation}
with
\begin{equation}
\mathcal{I}_{k}=4B\beta\sin^{2}{\left(\gamma-\frac{2}{3}\pi k\right)}
\end{equation}
and $\omega_{k}$ being the moments of inertia respectively the angular frequencies associated to the principal axes indexed by $k$. Due to the treatment of the vibrations and rotations on equal footing, the kinetic energy involves a single mass parameter $B$. A Schr\"{o}dinger equation in the $(\beta,\gamma,\Omega)$ coordinates is obtained by following the general prescription for quantization in curvilinear coordinates and the result is the well known Bohr-Mottelson Hamiltonian \cite{BM1,BM2}. However, the quantification procedure is based on the metric defined by the classical kinetic energy \cite{BM2}. Imposing some constraints on the kinetic energy one can reduce the number of variables and therefore obtain a different Hamiltonian. Indeed, by freezing the $\gamma$ variable ($\dot{\gamma}=0$) the vibrational energy becomes
\begin{equation}
T_{vibr}=\frac{B}{2}\dot{\beta}^{2}.
\label{kinv}
\end{equation}
In this case $\gamma$ is no longer a variable, being a simple parameter characterizing the shape. By quantizing now the system of remaining variables one recovers the Davydov-Chaban model \cite{Dav} for $\gamma$-rigid nuclei where the volume element is proportional to $\beta^{3}$ not to $\beta^{4}$ like in the usual Bohr-Mottelson approach.

Going further, and considering the axially symmetric prolate case ($\gamma=0$), one reduces the number of variables even more. For axially symmetric shapes the orientation angle with respect to the symmetry axis is indeterminate which leads to vanishing moment of inertia along the corresponding axis:
\begin{equation}
\mathcal{I}_{3}=0,\,\,\,\mathcal{I}_{1}=\mathcal{I}_{2}=3B\beta^{2}.
\end{equation}
In this situation, the rotational motion is described only by two degrees of freedom
\begin{equation}
T_{rot}=\frac{3}{2}B\beta^{2}\left(\omega_{1}^{2}+\omega_{2}^{2}\right)=\frac{3}{2}B\beta^{2}\left(\dot{\theta}_{1}^{2}\sin^{2}{\theta_{2}}+\dot{\theta}_{2}^{2}\right),
\end{equation}
while the vibrational motion described by (\ref{kinv}) is restricted only to oscillations preserving the axial symmetry.

The quantization in the cur\-vilinear coordinates $\theta_{1}$, $\theta_{2}$ and $\beta$ shape variable, provides the following operators for the two parts of the kinetic energy:
\begin{eqnarray}
T_{vibr}&=&-\frac{\hbar^{2}}{2B}\frac{1}{\beta^{2}}\frac{\partial}{\partial{\beta}}\beta^{2}\frac{\partial}{\partial{\beta}},\\
T_{rot}&=&\frac{\hat{\vec{I}}^{2}}{6B\beta^{2}},
\end{eqnarray}
where $\hat{\vec{I}}$ is the angular momentum in the intrinsic frame of reference. Now one can write the Hamiltonian associated to a prolate rigid nucleus as:
\begin{equation}
H=-\frac{\hbar^{2}}{2B}\left[\frac{1}{\beta^{2}}\frac{\partial}{\partial{\beta}}\beta^{2}\frac{\partial}{\partial{\beta}}-\frac{\hat{\vec{I}}^{2}}{3\hbar^{2}\beta^{2}}\right]+U(\beta).
\label{H}
\end{equation}
The Bohr-Mottelson Hamiltonian describes a variety of different types of collective motion depending on the potential energy function and the inertial parameters. The restrictions imposed so far changes not only the Hamiltonian but also its associated Hilbert space which in this case is defined by a metric proportional to $\beta^{2}$ consistent with the remaining three degrees of freedom, i.e. two Euler angles and the $\beta$ shape variable. It also leaves us only with the choice of the potential energy of the nuclear deformation $U(\beta)$, restricting thus the applicability of the above Hamiltonian only to the family of $\beta$ vibrational nuclei with $\gamma$-rigid axial symmetry. The Schr\"{o}dinger equation associated to such a Hamiltonian is solved by separating the $\beta$ variable from the angular ones which is achieved through the factorization:
\begin{equation}
\Psi(\beta,\theta_{1},\theta_{2})=F(\beta)Y_{IM}(\theta_{1},\theta_{2}),
\end{equation}
where the angular factor state is a spherical harmonic function and has the property:
\begin{equation}
\hat{\vec{I}}^{2}Y_{IM}(\theta_{1},\theta_{2})=I(I+1)\hbar^{2}Y_{IM}(\theta_{1},\theta_{2}).
\end{equation}
With this the Schr\"{o}dinger equation is reduced to a second order differential equation in variable $\beta$:
\begin{equation}
\left[\frac{1}{\beta^{2}}\frac{d}{d\beta}\beta^{2}\frac{d}{d\beta}-\frac{I(I+1)}{3\beta^{2}}+\frac{2B}{\hbar^{2}}\left(E-U(\beta)\right)\right]F(\beta)=0.
\end{equation}
This equation can be written also in the following form:
\begin{equation}
\left[\frac{d^{2}}{d\beta^{2}}+\frac{2}{\beta}\frac{d}{d\beta}-\frac{I(I+1)}{3\beta^{2}}+2(\varepsilon-u(\beta))\right]F(\beta)=0,
\label{BMR}
\end{equation}
where the following notations were used:
\begin{equation}
\varepsilon=\frac{B}{\hbar^{2}}E,\,\,\,\,u(\beta)=\frac{B}{\hbar^{2}}U(\beta).
\end{equation}
Up to this point the proceedings are exactly the same as in the construction of the $X(3)$ model \cite{X3}. As $X(3)$ is the $\gamma$ rigid version of the $X(5)$ critical point symmetry \cite{X5}, it uses an infinite square well potential which leads to a Bessel equation. Given the different choice of the potential, here one will treat equation (\ref{BMR}) differently from ref.\cite{X3}. Thus, making the change of function $F(\beta)=\frac{f(\beta)}{\beta}$ one obtains the equation:
\begin{equation}
\left[\frac{d^{2}}{d\beta^{2}}-\frac{I(I+1)}{3\beta^{2}}+2(\varepsilon-u(\beta))\right]f(\beta)=0,
\label{Fb}
\end{equation}
which resembles the radial Schr\"{o}dinger equation for a three-dimensional isotropic potential $u(\beta)$.

\renewcommand{\theequation}{3.\arabic{equation}}
\section{Quartic oscillator potential in $\beta$ variable}
\label{sec:2}
The quartic anharmonic oscillator potential in $\beta$ variable,
\begin{equation}
u(\beta)=\frac{1}{2}\alpha_{1}\beta^{2}+\alpha_{2}\beta^{4},
\label{Quartic}
\end{equation}
with $\alpha_{1},\alpha_{2}>0$, is the lowest order anharmonic potential when $\gamma=0$. With this choice of the potential $u(\beta)$ and the assumptions of the last section, the energy $\varepsilon$ then can be calculated as the eigenvalue of a Hamiltonian with a QAOP and a centrifugal barrier factorized by $I(I+1)/3$. The Schr\"{o}dinder equation for a QAOP cannot be solved exactly. However there are more than a few approximative methods for finding its eigenvalues which depend on the relative importance of the two terms. Here one will employ the method from ref.\cite{Quartic} for calculating the eigenvalues $\varepsilon$, which to our knowledge is the only one providing analytical formulas for it. Even though the prescription of ref.\cite{Quartic} refers to an $N$-dimensional quartic anharmonic oscillator it can be applied to the case of a more general centrifugal term given the fact that the orbital angular momentum is treated as a simple parameter in the derivation of the corresponding formulas. As the Eq.(\ref{Fb}) with an oscillator potential $u(\beta)=\beta^{2}$ can be brought to a form corresponding to Laguerre polynomials like in the case of the usual three-dimensional harmonic oscillator, the above mentioned procedure can indeed be applied in the present case because the numerical eigenvalues for QAOP are usually obtained in a harmonic oscillator basis. Although the formulas are derived on the basis of the forth order approximation made on the Jeffreys-Wentzel-Kramers-Brillouin (JWKB) quantization rule \cite{JWBK}, the numerical results are very precise with respect to extensive numerical computations \cite{num1,num2} for few selected potentials of the form (\ref{Quartic}). Moreover, the JWKB based energies in any order of approximation preserve the scaling property
\begin{equation}
\varepsilon(\alpha_{1},\alpha_{2})=\alpha_{2}^{1/3}\varepsilon(\alpha_{1}\alpha_{2}^{-2/3},1)=\alpha_{2}^{1/3}W(\alpha_{1}\alpha_{2}^{-2/3}),
\label{scal}
\end{equation}
of the exact eigenvalues. This means that one can find the eigenvalues of (\ref{Fb}) by solving the Schr\"{o}dinger equation for the potential
\begin{equation}
\tilde{u}(\beta')=\frac{1}{2}\lambda \beta'^{2}+\beta'^{4},
\label{ut}
\end{equation}
with $\lambda=\alpha_{1}\alpha_{2}^{-2/3}$. This is actually equivalent to the change of variable $\beta\,\rightarrow\,\beta'/\alpha_{2}^{1/6}$ in (\ref{Fb}). Following the procedure of ref.\cite{Quartic} one can write the following formula for the energy value $W$:
\begin{equation}
W_{nI}(\lambda)=\left(N_{nI}\right)^{\frac{4}{3}}\sum_{k=0}^{8}G_{k}(\lambda,I)\left(N_{nI}\right)^{-\frac{2k}{3}},
\label{W}
\end{equation}
where
\begin{equation}
N_{nI}=\sqrt{\frac{9\pi\eta}{8}}\left[2n+1+\frac{1}{2}\sqrt{1+\frac{4I(I+1)}{3}}\right],
\end{equation}
with $\eta$ being a constant which is given together with the functions $G_{k}(\lambda,I)$ in Appendix. The energy (\ref{W}) is indexed by the principal quantum number $n$ which is associated to the number of $\beta$ vibration quanta and the angular momentum $I$ of the collective rotation. The dependence on angular momentum $I$ comes from the angular degrees of freedom defining the centrifugal term.

The energy of a nucleus described by the rotation-vibration Hamiltonian (\ref{H}) with QAOP is then given by
\begin{equation}
E_{nI}=\frac{\hbar^{2}}{B}\alpha_{2}^{1/3}\left[W_{nI}(\lambda)-W_{00}(\lambda)\right].
\label{En}
\end{equation}
So up to an overall scaling factor the energies depend only on one parameter, namely $\lambda$. When $\lambda=0$ the problem is reduced to the pure quartic oscillator, where the above analytical formula for the energy levels still holds but no longer depend on any parameter except the factorizing one. In this case one obtains a parameter free model which will be called $X(3)$-$\beta^{4}$ on account of the same structure as the already established $X(3)$ model. Considering the other limit of the quartic potential, i.e. $\lambda\rightarrow\infty$, one recovers the harmonic oscillator case. It must be emphasized that the adopted formalism does not yield the correct harmonic oscillator energy levels because the expansion of the JWKB integrals is about the pure quartic oscillator levels. However, obtaining the harmonic oscillator results by solving eq.(\ref{Fb}) with an oscillator potential $u(\beta)=\beta^{2}$ is straightforward, providing the following energy
\begin{equation}
E_{nI}^{ho}=\frac{\hbar^{2}}{B}\left[2n+\frac{1}{2}\left(\sqrt{1+\frac{4I(I+1)}{3}}-1\right)\right],
\label{Eho}
\end{equation}
normalized to the ground state. In this way one produced another parameter free model in the same three-\-dimension\-al space which will be denoted hereafter as $X(3)$-$\beta^{2}$, following the above mentioned reasons.

An important consequence of the analytical expression of the QAOP energy is the possibility to derive the expectation values for the even powers of the associated radial variable from hypervirial relations \cite{Hughes} using Hellmann-Feynman theorem. Thus, taking into account the scaling relation (\ref{scal}) and the prescription of ref.\cite{Quartic} one can express the first order even moment as follows:
\begin{equation}
\langle n,I| \beta^{2}|n,I\rangle=\alpha_{2}^{-1/3}\frac{\partial{W_{nI}}}{\partial{\lambda}}.
\label{b2}
\end{equation}
Unfortunately, except some recurrence relations between different order moments \cite{Hughes}, there are no analytical formulas available for the nondiagonal moments $\beta^{2k}$ which would be useful for calculating electromagnetic transition probabilities.

\section{Model applicability}
\label{sec:3}

The energy levels of the ground state band ($n=0$), as well as of the $\beta$ vibrational bands ($n>0$) defined by equation (\ref{En}) and normalized to the energy of the lowest excited state $2^{+}_{1}$ depend only on $\lambda$. The shapes of the axially symmetric nuclei are parametrized by a nuclear deformation parameter which is not a directly measurable observable. Then instead one usually describes different nuclear shape phases in terms of the ratio $R_{4/2}$ between the lowest two collective energy levels $4^{+}_{1}$ and $2^{+}_{1}$. In the framework of IBM \cite{IBM} to each shape phase there is associated a dynamical symmetry whose signature is a specific value of $R_{4/2}$. Thus it should be useful to check the applicability of our model by studying the dependence of $R_{4/2}$ on the parameter $\lambda$. This is done in Fig. 1, where one also represented the ratio between the second $0^{+}_{2}$ state energy from the first $\beta$ excited band and the $2^{+}_{1}$ ground band energy. Beside $R_{4/2}$, the ratio $E(0^{+}_{2})/E(2^{+}_{1})$ is another key signature characterizing the dynamical symmetries, especially those associated to a critical point of a shape phase transition. The specific polynomial structure of the energy (\ref{W}) as function of $\lambda$ visualized in Fig.\ref{W0I} produce a minimum at $\lambda_{min}=15.0649$ and a pole in $R_{4/2}$ at the value $\lambda_{p}=17.4323$ where $W_{00}$ and $W_{02}$ intersect each other. It becomes clear from Fig.\ref{W0I} that the adopted method for the determination of the QAOP eigenvalues provides reliable results only for a limited interval of $\lambda$.

\begin{figure}
\resizebox{0.47\textwidth}{!}{%
\includegraphics{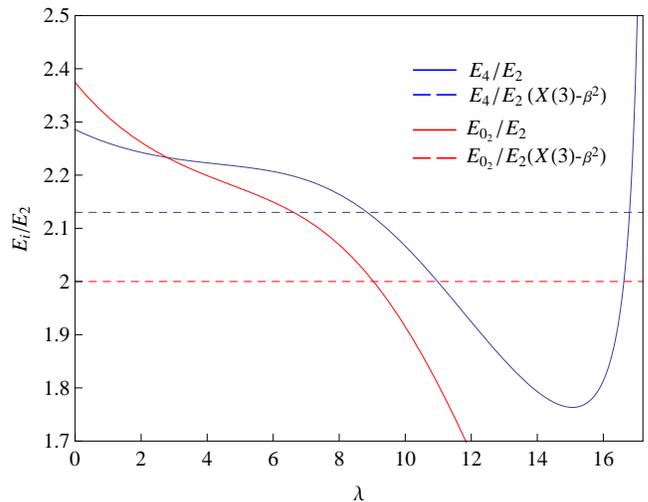}}
\caption{The theoretical ratios $R_{4/2}=E(4^{+}_{1})/E(2^{+}_{1})$ and $E(0^{+}_{2})/E(2^{+}_{1})$ provided in the framework of QAOP are given as functions of $\lambda$. For reference, the same ratios corresponding to the $X(3)$-$\beta^{2}$ model are also visualized.}
\label{4p2}
\end{figure}

\begin{figure}
\resizebox{0.47\textwidth}{!}{%
\includegraphics{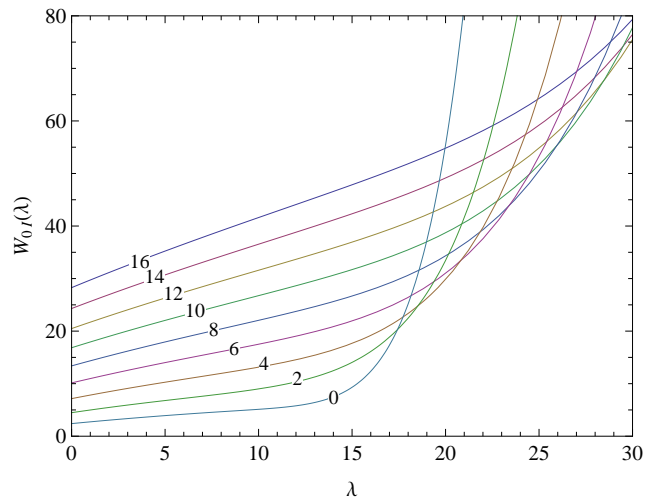}}
\caption{The scaled energy $W$ (\ref{W}) given as function of $\lambda$ for $n=0$ and different angular momenta $I$.}
\label{W0I}
\end{figure}

It is well known that the accuracy of the JWKB generated eigenvalues is increasing with the order of the solution \cite{Hioe}, i.e. the quantum numbers. Thus, in order to establish the upper limit of the parameter $\lambda$ for which one would still obtain reliable eigenvalues, it is sufficient to compare the present numerical results with the available exact eigenvalues only for the ground state. Taking as a benchmark the exact ground state eigenvalue computed in ref.\cite{num1} for the three-dimensional potential $u(r)=\frac{1}{2}r^{2}+0.025\,r^{4}$ which corresponds to $\lambda=11.6961$, one finds that the present calculations overestimates the corresponding eigenvalue by 4.3\% which is more than satisfactory. However, in the view of the present physical context, one will choose for the upper limit of $\lambda$ the value $\lambda_{cutoff}=10.9802$ which is associated to the ratio $R_{4/2}=2$ characterizing vibrational nuclei, ensuring in this way an even better accuracy. Thus, in the following calculations one will consider the values of $\lambda$ only from the interval $[0,\lambda_{cutoff}]$ where $R_{4/2}$ is unambiguously defined with values ranging continuously from 2.286 to 2, and correspondingly with $E(0^{+}_{2})/E(2^{+}_{1})$ taking values between 2.374 and 1.808 as is shown in Fig. 1. In the same figure one also visualised for comparison the ratios provided by $X(3)$-$\beta^{2}$ model, which amounts to $R_{4/2}=2.13$ and $E(0^{+}_{2})/E(2^{+}_{1})=2$, consequently falling in the existence interval of the present model.

\begin{figure}
\resizebox{0.47\textwidth}{!}{%
\includegraphics{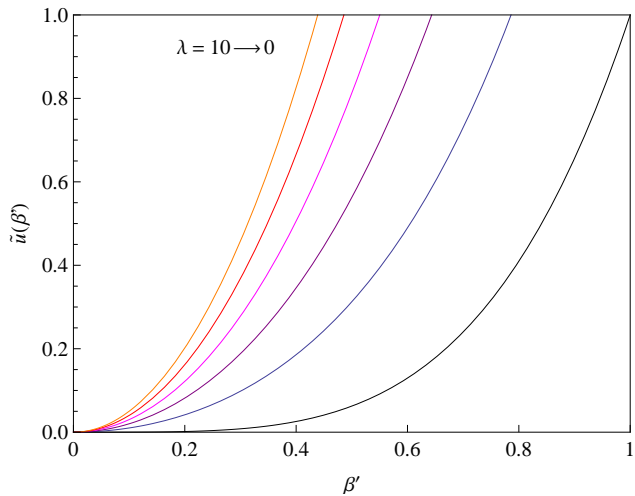}
}
\caption{The scaled potential (\ref{ut}) given as function of $\beta'$ for different values of $\lambda$ ranging from 0 to 10 with step 2.}
\label{Pot}
\end{figure}

Although the potential (\ref{Quartic}) depends on two parameters, $\alpha_{1}$ and $\alpha_{2}$, the normalized energy spectrum can be described only by $\lambda$. As a matter of fact, the shape of the potential (\ref{Quartic}) is also determined only by $\lambda$ through its scaled version (\ref{ut}). Indeed, the scaled potential (\ref{ut}) is a function of $\beta'$ variable which depend on $\alpha_{2}$, such that for the same $\lambda$ the potential (\ref{Quartic}) will have the same shape with a larger width for smaller $\alpha_{2}$. The shapes of the scaled potential for different values of $\lambda$ are shown in Fig.\ref{Pot} as function of the scaled variable $\beta'$. From this figure one can see that the flattest potential corresponds to the pure quartic oscillator case $\lambda=0$. As the critical point potential must exhibit a flat behaviour, it can be inferred that the pure quartic oscillator potential is a critical one at least for the family of the general anharmonic potentials of the form (\ref{Quartic}) in the $\gamma$-rigid regime.

\begin{figure}
\resizebox{0.47\textwidth}{!}{%
\includegraphics{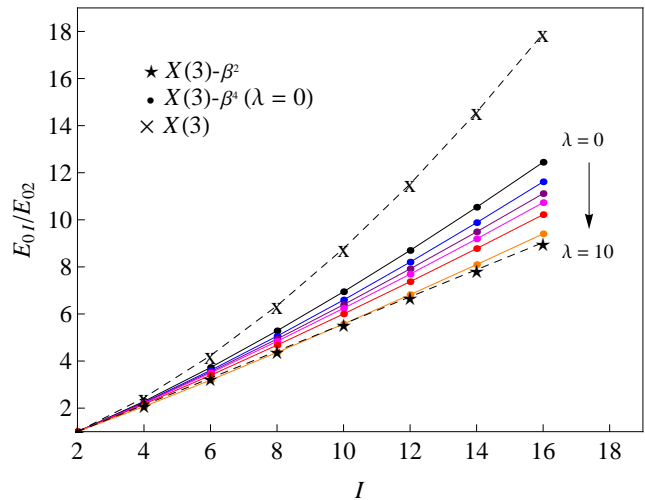}}
\caption{Ground band energy normalized to the first excited state energy given as function of angular momentum for different values of parameter $\lambda$ ranging from 0 to 10 with step 2. The $X(3)$ predictions taken from ref.\cite{X3} and the $X(3)$-$\beta^{2}$ results obtained with (\ref{Eho}) are also shown for guidance.}
\label{EJ}
\end{figure}

\begin{figure}
\resizebox{0.47\textwidth}{!}{%
\includegraphics{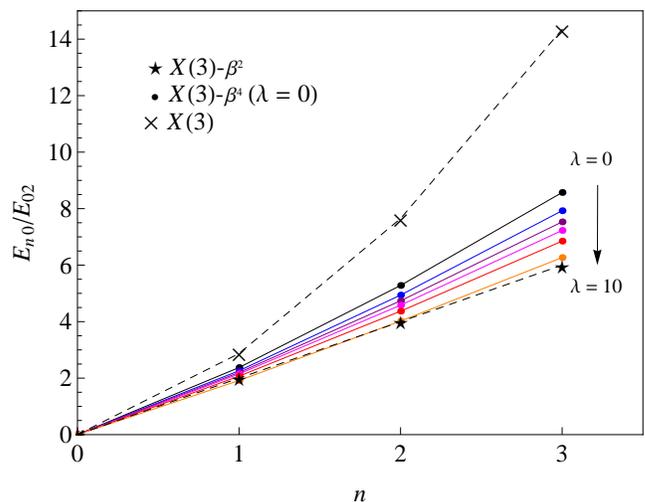}}
\caption{Bandheads energies normalized to the first excited state energy given as function of quantum number $n$ indexing the band for different values of parameter $\lambda$ ranging from 0 to 10 with step 2. The $X(3)$ predictions taken from ref.\cite{X3} and the $X(3)$-$\beta^{2}$ results obtained with (\ref{Eho}) are also shown for guidance. The quantum number $s$ from ref.\cite{X3} corresponds to $n+1$ in the present indexing of bands.}
\label{Enn}
\end{figure}

\begin{figure*}[th!]
\resizebox{1\textwidth}{!}{%
\includegraphics[trim = 5mm 0mm 5mm 0mm, clip]{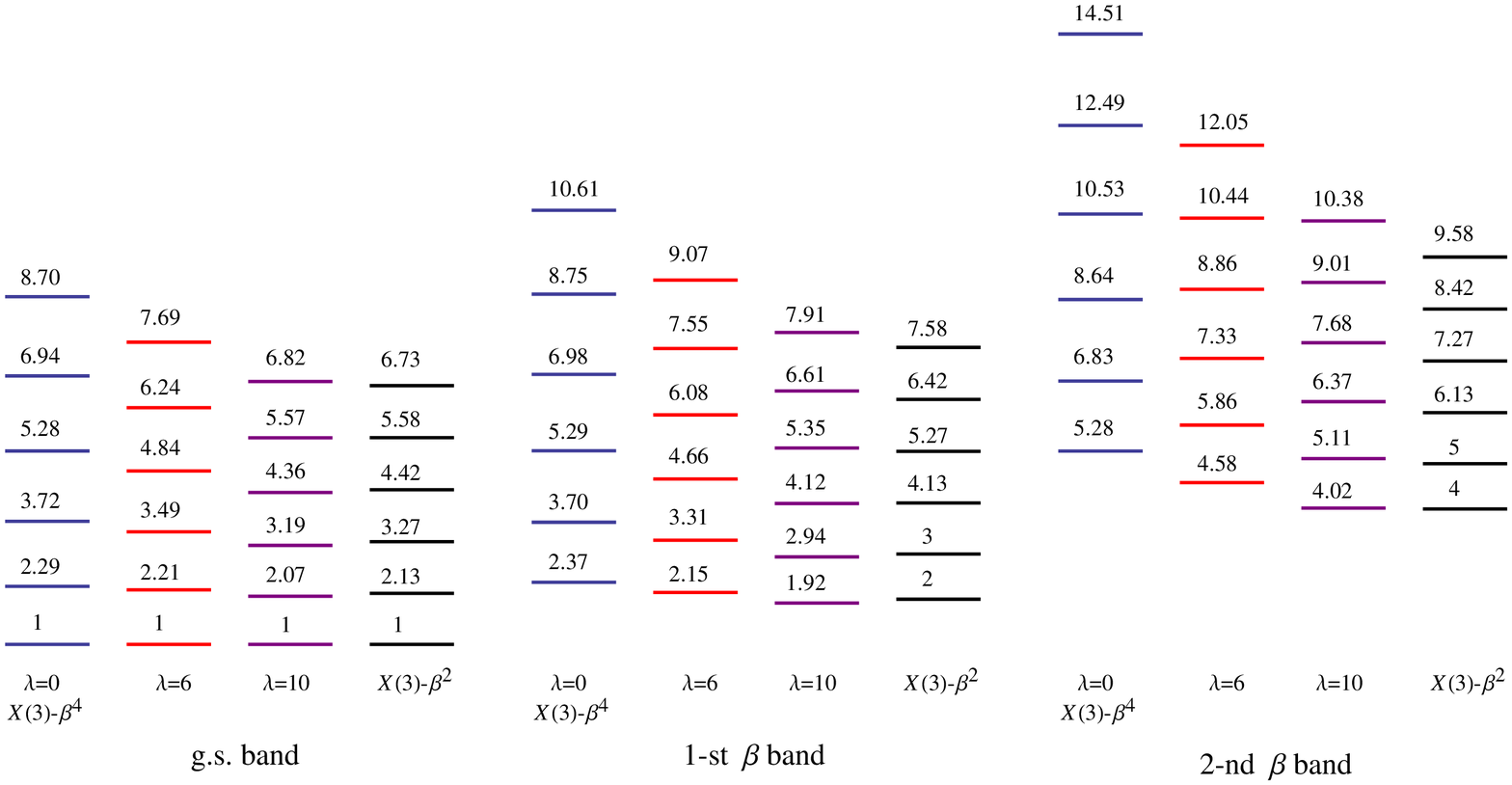}}
\caption{The theoretical ground, first and second excited $\beta$ bands spectra for $\lambda=0,6$ and 10 compared to the corresponding predictions of the $X(3)$-$\beta^{2}$ model.}
\label{Spectra}
\end{figure*}

Concerning the allowed values only for $R_{4/2}$, it should be noted that the highest value $2.286$ is achieved for the pure quartic potential, while the lowest admissible value $R_{4/2}^{min}=2.0$ is very close to the signature of the $X(3)$-$\beta^{2}$ model described by $R_{4/2}=2.13$. Moreover, the flexible structure of the present model allows the description of all interpolating solutions lying in between the $X(3)$-$\beta^{2}$ and $X(3)$-$\beta^{4}$ model realizations through the continuous variation of a single parameter, $\lambda$. This transition is best seen in the evolution of the ground band as function of angular momentum for values of $\lambda$ corresponding to the shapes of the scaled potential depicted in Fig.\ref{Pot} which is visualized in Fig.\ref{EJ} where one also shown the $X(3)$ \cite{X3} and $X(3)$-$\beta^{2}$ predictions. One can easily observe that for $\lambda=0$ the ground band spectrum has a behaviour distinct from those corresponding to $\lambda\neq0$, being the steepest one. As a matter of fact, the smaller the value of $\lambda$, more closer to the rotational behaviour $I(I+1)$ is the corresponding ground band spectrum and consequently farther from the vibrational behaviour $\sim I$. Moreover, the $\lambda=0$ case seems to be positioned at the half way between the $X(3)$ and $X(3)$-$\beta^{2}$ model predictions with the lower region covered by the present model with $\lambda>0$. In what concerns the spectra situated between $X(3)$ and $X(3)$-$\beta^{4}$, these can be described by eventual $X(3)$-$\beta^{2n}(n>2)$ critical models as in the five-dimensional phase transitions \cite{X5b,Bon2004}. The same picture is also found in the dependence of the energy spectrum on the vibrational quanta $n$ for $I=0$ shown in Fig.\ref{Enn}. Another interesting feature arising from Figs.\ref{EJ} and \ref{Enn} is that the spectra corresponding to values of $\lambda$ up to approximatively 8 are somehow bunched together and more closely at low angular momentum states. This phenomenon can be ascribed to the fact that at the same value the curve of $R_{4/2}$ from Fig.\ref{4p2} has a significant change in its tangent. Making a more detailed analysis of the function $R_{4/2}(\lambda)$ one ascertains that its third derivative vanishes at $\lambda=8.75$ where $R_{4/2}=2.135$. The mathematical meaning of this fact is that the angle between the axis of the osculating parabola and the normal line associated to that point is equal to the angle made by the tangent in that point. Although not in the true sense of the theory of the quantum phase transitions \cite{Cejnar}, one can say that there is a transition between two weakly delimited phases defined by the two sides of the value $\lambda=8.75$. Combining the Figs.\ref{EJ} and \ref{Enn}, one visualized in Fig.\ref{Spectra} the ground band up to $I=12$ together with the first and second $\beta$ bands spectra up to $I=10$ for the pure quartic case ($\lambda=0$) and for two values of $\lambda$ separated by the "critical" point $\lambda_{c}=8.75$ followed by the corresponding $X(3)$-$\beta^{2}$ results. It is worth to mention that even thought the present formalism cannot be extended to the harmonic oscillator case, the $X(3)$-$\beta^{2}$ predictions are fairly well simulated by our model with $\lambda>\lambda_{c}$ at least for the lowest lying states.

\section{Numerical application}
\label{sec:4}

\setlength{\tabcolsep}{10.85pt}
\begin{table*}[th!]
\caption{The values of $\lambda$ obtained from experimental $R_{4/2}$ ratios are given for each treated nucleus together with the corresponding parameters $\alpha_{1}$ and $\alpha_{2}$ of the initial potential (\ref{Quartic}) extracted from (\ref{b2}) with the tabulated $\beta_{2}$ value used in the left hand side. The values of $\beta_{2}$ were taken from \cite{Lala}.}
\label{tab:2}
\begin{center}
\begin{tabular}{rcrrrrr}
\hline\noalign{\smallskip}
Nucleus &$R_{4/2}$& $\lambda$ & $\beta_{2}$& $\frac{\partial{W_{00}}}{\partial{\lambda}}$ &$\alpha_{1}$ & $\alpha_{2}$ \\
\noalign{\smallskip}\hline\noalign{\smallskip}
$^{100}$Mo &2.121& 9.031 & 0.253&0.449&            345.13&            444.35\\
$^{100}$Pd &2.128& 8.893 & 0.136&0.447&          14157.54&           5204.48\\
$^{116}$Te &2.002&10.949 & 0.257&0.626&            853.21&            984.91\\
$^{130}$Xe &2.247& 1.705 & 0.128&0.631&          57066.90&           2526.61\\
$^{148}$Sm &2.145& 8.508 & 0.112&0.447&          45313.50&          10814.20\\
$^{152}$Gd &2.194& 6.834 & 0.178&0.483&           3543.80&           1588.56\\
$^{154}$Dy &2.234& 2.739 & 0.179&0.598&           6516.71&            955.75\\
$^{154}$Er &2.072& 9.910 & 0.147&0.485&          11333.37&           5000.14\\
$^{220}$Th &2.035&10.479 & 0.012&0.545&   $5.42\,10^{10}$&    1.50$\,10^{8}$\\
\noalign{\smallskip}\hline
\end{tabular}
\end{center}
\end{table*}

\begin{figure*}[th!]
\resizebox{1\textwidth}{!}{%
\includegraphics[trim = 0mm 0mm 0mm 0mm, clip,width=1\textwidth]{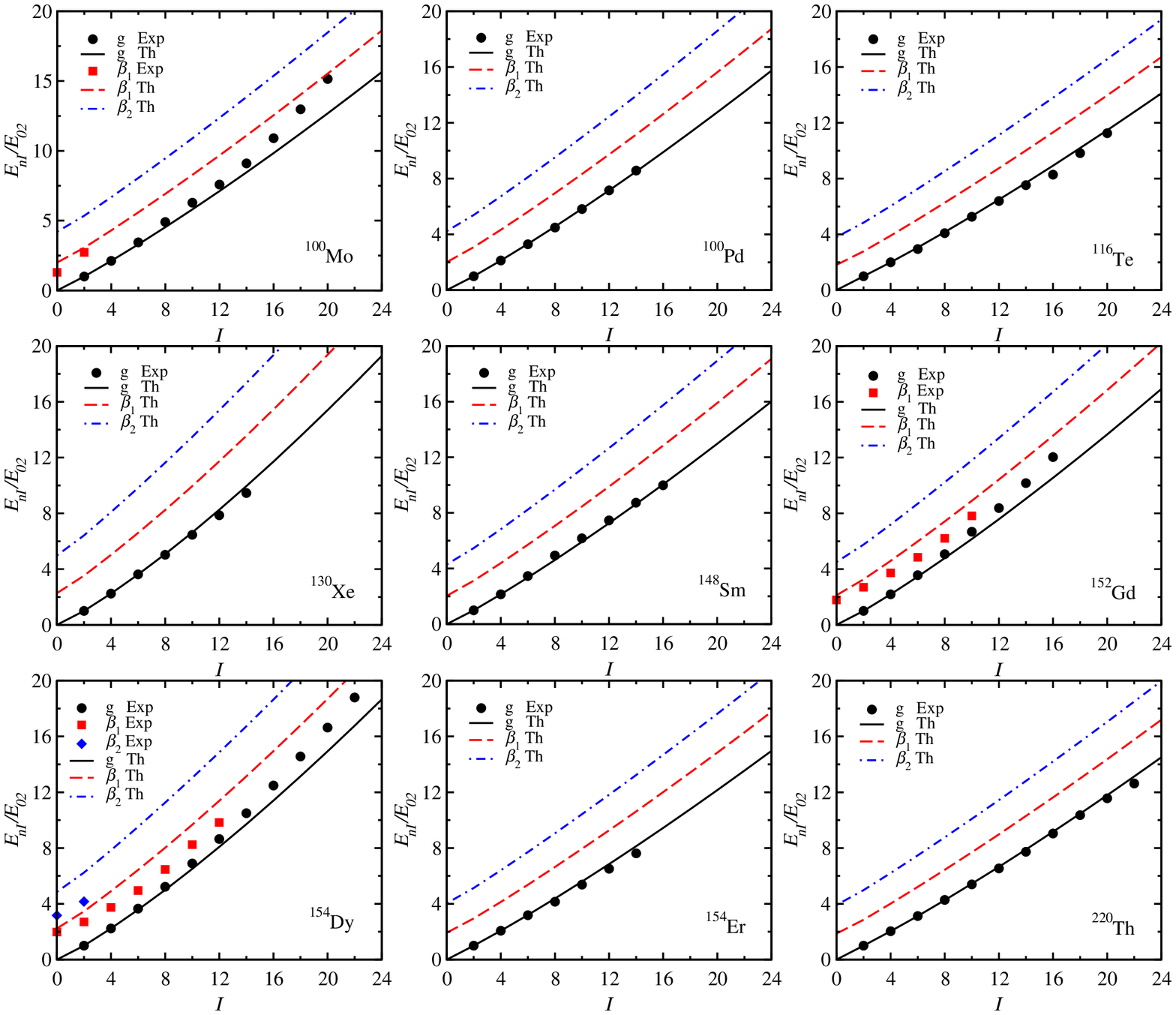}}
\caption{Theoretical results for ground, first and second excited $\beta$ bands energies normalized to the energy of the $2^{+}$ ground state are compared with the available experimental data for $^{100}$Mo \cite{MoPd100}, $^{100}$Pd \cite{MoPd100}, $^{116}$Te \cite{Te116}, $^{130}$Xe \cite{Xe130}, $^{148}$Sm \cite{Sm148}, $^{152}$Gd \cite{Gd152}, $^{154}$Dy \cite{DyEr154}, $^{154}$Er \cite{DyEr154} and $^{220}$Th \cite{Th220}.}
\label{ExpTh}
\end{figure*}

In order to see how the model presented in the last sections behave itself when applied to concrete nuclei, one browsed the nuclide chart in search of nuclei with collective spectrum populated with at least seven states and whose experimental ratio $R_{4/2}$ falls in the existence interval of our model. There were found a lot of nuclei which satisfied the above mentioned criteria, but only to a handful of them the present model could be successfully applied, namely $^{100}$Mo, $^{100}$Pd, $^{116}$Te, $^{130}$Xe, few lighter rare earth isotopes $^{148}$Sm, $^{152}$Gd, $^{154}$Dy, $^{154}$Er, and a trans-lead nucleus $^{220}$Th. All these nuclei have $R_{4/2}>2$, as it is expected because a value $R_{4/2}<2$ characterizes nuclei in the vicinity of a shell closure where single-particle degrees of freedom prevail. For each of these nine nuclei one determined the parameter $\lambda$ which is listed in Table \ref{tab:2} by equating the theoretical and experimental ratio $R_{4/2}$. The calculated values for $\lambda$ are then used to calculate higher angular momentum state energies of the ground and $\beta$ excited bands. The numerical results obtained in this way for the ground, first and second $\beta$ excited bands energies normalized to the energy of the $2_{1}^{+}$ state are compared with corresponding available experimental data in Fig.\ref{ExpTh}. Only three of the considered nuclei, $^{100}$Mo, $^{152}$Gd and $^{154}$Dy, also have few experimentally observed states in the $\beta$-vibrational bands, while for nuclei $^{148}$Sm and $^{154}$Er one considered only the ground band states up to the clearly visible backbending happening between $I=14$ and $I=16$.

The original form of the potential (\ref{Quartic}), i.e. the parameters $\alpha_{1}$ and $\alpha_{2}$, can be recovered from equation (\ref{b2}) and thus being comparable with other choices for $\beta$ potential (square well, harmonic oscillator, etc.). Indeed, by equating the square of the ground state deformation $\beta_{2}$ \cite{Lala} with the theoretical value (\ref{b2}) corresponding to $\lambda$ calculated above, one would obtain an equation for determining $\alpha_{2}$. The remaining parameter $\alpha_{1}$ is then recovered from the definition of $\lambda$. The numerical values of $\alpha_{1}$ and $\alpha_{2}$ defining the potential (\ref{Quartic}) are given in Table \ref{tab:2} for each treated nucleus, where one also calculated the quantity $\partial{W_{00}}/\partial{\lambda}$ which is essential in determining $\alpha_{2}$.

An overall impression of the comparison between the theoretical and experimental spectra presented in Fig. \ref{ExpTh} is that the theoretical predictions keep up with the corresponding experimental values for few lower states, and after there is a regression in the agreement between them. This is due to the fact that one fixed the free parameter $\lambda$ by fitting the ratio $R_{4/2}$, rather than fitting the whole spectrum which might provide a better agreement for higher spin states. The procedure adopted here for fixing $\lambda$ is justified by its direct relation to an important observable, $R_{4/2}$. Regardless, the agreement with experiment over the whole spectrum is quite good for all considered nuclei. A special attention is deserved by $^{100}$Pd which is well known as the most promising candidate for the $E(5)$-$\beta^{4}$ model \cite{Bon2004,Vorov}. As it happens, the experimental spectrum limited only to the ground band, is best reproduced for this nucleus. The prediction of $E(5)$-$\beta^{4}$ model \cite{Bon2004,Vorov} and those of the present formalism for this nucleus are equally good, alternating the best agreement at different states, even though the structure and the acting space of the two models are completely different. New experimental measurements regarding the collective states of this nucleus will eventually incline the scales toward the more suitable description.

Checking the nuclear deformation listed in Table \ref{tab:2} for the treated nuclei, one can observe that it ranges from very small $\beta_{2}=0.012$ for $^{220}$Th to considerably large $\beta_{2}=0.257$ for $^{116}$Te even though all nuclei have a vibrational-like collective spectrum. Apart from $^{100}$Pd, the two extremes in terms of the nuclear deformation, $^{116}$Te and $^{220}$Th, together with the rare earth nucleus $^{148}$Sm, are in the view of the agreement with experiment the best representatives of the model introduced in this paper. Although with a poorer reproduction of the experimental spectrum, the nuclei $^{100}$Mo, $^{152}$Gd and $^{154}$Dy are also promising cases due to the simultaneous description of the $\beta$ excited band states which are fairly well reproduced considering that the model has a single free parameter. The addition of experimental data for $\beta$ excited states for the rest of the considered nuclei will be an important test of the present model as its quality is given at this moment only by the ground band spectrum.

Before closing this section it is worth to mention the fact that the treated nuclei happen to be separated almost equally with respect to the critical value $\lambda_{c}=8.75$. The experimental energy spectrum of the nuclei with $\lambda>8.75$ seem to be better reproduced by the theoretical calculations than the other half. Exception is the $^{148}$Sm nucleus whose corresponding value of $\lambda$ is not much smaller than the critical value.

\section{Conclusions}
\label{sec:5}
An analytical formula for the energies of the ground and $\beta$ vibrational bands was derived in the framework of the prolate $\gamma$-rigid regime of the Bohr-Mottelson Hamiltonian with a quartic oscillator potential in $\beta$ shape variable. The differential equation in $\beta$ for a QAOP is not exactly solvable, such that the formula proposed is based on higher order JWKB approximation. The energy formula depend on a single free parameter up to an overall multiplying constant. Studying the convergence of the adopted approximation to the exact results one fixed the upper limit of the free parameter at $\lambda_{cutoff}=10.9802$ which corresponds to $R_{4/2}=2$. The model applicability is then established by the dependence of the $R_{4/2}$ ratio on the free parameter $\lambda$ restricted to the interval $[0,\lambda_{cutoff}]$. For $\lambda=0$, which marks the $X(3)$-$\beta^{4}$ model, one obtains the maximum value of the $R_{4/2}$ ratio, which is 2.286. On the other hand, the model predictions in the vicinity of $\lambda_{cutoff}$ simulate quite well the spectrum of the $X(3)$-$\beta^{2}$ model. In virtue of these limiting cases as well as of the results for various values of $\lambda$ (see Figs. \ref{EJ} and \ref{Enn}) one can say that the proposed model represents a bridge between $X(3)$-$\beta^{2}$ and $X(3)$-$\beta^{4}$ models with a complete set of interpolating solutions defined by the continuous variation of the parameter $\lambda$. In addition, studying the behaviour of the ground band energy spectrum and the evolution of the bandheads indexed by vibrational quanta $n$, for different values of the free parameter $\lambda$, one identified a turning point at $\lambda_{c}=8.75$ which separates two "phases" characterized by specific features of the collective spectrum.

The model was successfully applied for nine nuclei covering different parts of the nuclide chart. Indeed, even by fixing the free parameter to reproduce the experimental $R_{4/2}$ ratio, a qualitative reproduction of the whole experimental ground band spectrum is obtained for all nuclei and even of the $\beta$ excited bands when experimentally available but with less precision. Moreover, the present formalism produce almost the same agreement with experiment for $^{100}$Pd as the $E(5)$-$\beta^{4}$ model even though the two span different spaces of shape and angular degrees of freedom. It is worth to mention that the best agreement with experiment is obtained for the nuclei with $\lambda$ above the critical value 8.75.

A few perspectives of the present approach are to be pinpointed before closing. The QAOP can be further used to investigate $\gamma$-rigid nuclei with $\gamma\neq0$ as in the $Z(4)$ \cite{Z4} model of Bonatsos or to generate more extensive predictions regarding the energy spectrum in the $\gamma$-unstable case of the collective geometrical model. Another possible application refers to the other branch of the symmetries' triangle where the separation of the shape variables is only approximative. Concluding, the novelty of the present formalism consists in the introduction of an analytical formula for the energy which was shown to be able to describe nuclei with irregular vibrational-like spectra.

\section*{Appendix}
\label{App}

The functions $G_{k}(\lambda,I)$ defining the energy $W_{nI}(\lambda)$ are taken from \cite{Quartic} and adjusted to the present physical problem acquire the following expressions:
\begin{eqnarray*}
G_{0}&=&1,\,\,\,\,G_{1}=\frac{\lambda\eta}{2},\,\,\,\,G_{2}=-\frac{\lambda^{2}}{32}(1-3\eta^{2}),
\end{eqnarray*}
\begin{eqnarray*}
G_{3}&=&\frac{\eta}{24}\left[3+\frac{1}{8}\lambda^{3}\eta^{2}-4I(I+1)\right],\\
G_{4}&=&\frac{\lambda}{192}\left[1-3\eta^{2}+\frac{1}{64}\lambda(1-5\eta^{4})+4(1+\eta^{2})I(I+1)\right],\\
G_{5}&=&-\frac{\lambda^{2}\eta}{1280}\left[5+\frac{1}{16}\lambda^{3}+20I(I+1)\right],\\
G_{6}&=&\frac{1}{192}\left[-\frac{11}{8}-\frac{15}{8}\eta^{2}+\frac{5}{64}\lambda^{3}(-1+4\eta^{2}+\eta^{4})\right.\\
&&+\frac{7}{6144}\lambda^{6}\eta^{2}(3+\eta^{4})+\left[-25+15\eta^{2}\right.\\
&&\left.+\frac{1}{16}\lambda^{3}(5+60\eta^{2}-5\eta^{4})\right]\frac{I(I+1)}{3}\\
&&+(10-30\eta^{2})\frac{I^{2}(I+1)^{2}}{9}\Big],\\
G_{7}&=&\frac{7\lambda\eta}{768}\left[-\frac{39}{4}+\frac{3}{4}\eta^{2}+\frac{1}{64}\lambda^{3}\left(\frac{7}{2}-\frac{10}{3}\eta^{2}-\frac{1}{2}\eta^{4}\right)\right.\\
&&-\frac{1}{1536}\lambda^{6}\eta^{2}\left(\frac{4}{5}+\frac{1}{7}\eta^{4}\right)\\
&&+\left[28-6\eta^{4}+\frac{1}{32}\lambda^{3}(-7-20\eta^{2}+\eta^{4})\right]\frac{I(I+1)}{3}\\
&&\left.+(-8+12\eta^{2})\frac{I^{2}(I+1)^{2}}{9}\right],\\
G_{8}&=&\frac{9\lambda^{2}}{4096}\left[\frac{23}{24}+\frac{95}{2}\eta^{2}-\frac{9}{8}\eta^{4}+\frac{1}{64}\lambda^{3}\left(\frac{1}{9}-9\eta^{2}+3\eta^{4}\right)\right.\\
&&+\frac{1}{4096}\lambda^{6}\left(\frac{1}{63}+\frac{9}{5}\eta^{4}\right)+\left[-\frac{31}{3}-96\eta^{2}+9\eta^{4}\right.\\
&&\left.+\frac{1}{64}\lambda^{3}\left(\frac{4}{3}+\frac{192}{5}\eta^{2}+36\eta^{4}\right)\right]\frac{I(I+1)}{3}\\
&&\left.+\left(\frac{14}{3}+16\eta^{2}-18\eta^{4}\right)\frac{I^{2}(I+1)^{2}}{9}\right],
\end{eqnarray*}
where  $\eta=\left[\frac{2\Gamma\left(\frac{3}{4}\right)}{\Gamma\left(\frac{1}{4}\right)}\right]^{2}=0.457.$

%
%

\end{document}